\documentstyle[aps,prl,preprint,floats,epsf,psfig]{revtex}

\begin{document}

\preprint{\tighten\vbox{\hbox{\hfil CLNS 97/1469}
                        \hbox{\hfil CLEO 97-4}
}}

\title{\bf Search for neutrinoless $\tau$ decays
involving $\pi^0$ or $\eta$ mesons}

\author{CLEO Collaboration}
\date{\today}

\maketitle
\tighten

\begin{abstract}
We have searched for 
lepton family number violating decays of the $\tau$
lepton using final states with an electron or a muon and
one or more $\pi^0$ or $\eta$ mesons but no neutrinos.
The data used in the search were collected with the
CLEO II detector at the Cornell Electron Storage Ring (CESR) and
correspond to an integrated luminosity of 4.68 $fb^{-1}$.
No evidence for signals was found, resulting in much improved
limits on the branching fractions for the one-meson modes
and the first upper limits for the
two-meson modes. 
\end{abstract}
\newpage

{
\renewcommand{\thefootnote}{\fnsymbol{footnote}}

\begin{center}
G.~Bonvicini,$^{1}$ D.~Cinabro,$^{1}$ R.~Greene,$^{1}$
L.~P.~Perera,$^{1}$ G.~J.~Zhou,$^{1}$
B.~Barish,$^{2}$ M.~Chadha,$^{2}$ S.~Chan,$^{2}$ G.~Eigen,$^{2}$
J.~S.~Miller,$^{2}$
% ...
\end{center}
 
\small
\begin{center}
$^{1}${Wayne State University, Detroit, Michigan 48202}\\
$^{2}${California Institute of Technology, Pasadena, California 91125}\\
% ...
\end{center}

\setcounter{footnote}{0}
}
\newpage

In physics all fundamental conservation laws have 
associated symmetries.  Lepton flavor conservation is an experimentally
observed phenomenon with no associated symmetry in the Standard Model.
Lepton flavor violation is expected in many extensions of
the Standard Model such as lepto-quark, supersymmetry, superstring,
and left-right symmetric models, and models that
include heavy neutral leptons.
Gonzalez-Garcia and Valle have calculated~\cite{Valle}, in a model with
Dirac heavy neutral leptons, the branching fractions for $\tau$ decay into one lepton plus
a photon or a $\pi^0$ or $\eta$ meson.
The branching fractions depend on the heavy neutral lepton masses and mixings.
Given the constraints from other measurements, the branching fraction for
$\tau^- \to \ell^- \pi^0$~\cite{light} may still be as large as $10^{-6}$ 
for neutral lepton
masses above a few TeV/c$^2$ and is higher than that for
the radiative decay $\tau^- \to \ell^- \gamma$.
Using a Grand Unified Theory (GUT) and superstring inspired model with heavy neutral leptons,
Ilakovoc and collaborators~\cite{Ilakovoc} have calculated
the branching fractions for $\tau$ decay into one lepton plus one or two mesons.
The branching fractions depend on the masses of the Majorana neutrinos and
the mixings between the heavy and light neutrinos.
The decay $\tau^- \to \ell^- \pi^0$ may have a branching fraction
as large as $10^{-6}$.
The previous upper limits~\cite{pdg} on the branching fractions
for the decays into one lepton and a $\pi^0$ or $\eta$ meson are of the order
of $10^{-4}-10^{-5}$ and for the decays into one lepton and two charged $\pi$
mesons are of the order of $10^{-6}$.
There are no published results for the decays into one lepton and two neutral mesons 
($\pi^0 \pi^0$, $\eta\eta$ or $\pi^0\eta$).
The CLEO~II experiment with its large sample of $\tau$ events 
may have the sensitivity to observe the lepton flavor violating decays.
In this Letter, we present the result of a search for the decays into one lepton and one
or two $\pi^0$ or $\eta$ mesons.

The data used in this analysis were collected with the CLEO~II
detector from $e^+e^-$ collisions at the Cornell Electron
Storage Ring (CESR) at a center-of-mass energy $\sqrt{s} \sim 10.6$ GeV.
The total integrated luminosity of the data sample is 4.68 fb$^{-1}$,
corresponding to the production of
$N_{\tau\tau} = 4.26 \times 10^6\ \tau^+\tau^-$ events.
CLEO~II is a general purpose spectrometer~\cite{cleoiinim} with
excellent charged particle and shower energy detection.
The momenta and specific ionization
(dE/dx) of charged particles are measured with three
cylindrical drift chambers between 5 and 90~cm from the
$e^+e^-$ interaction point, with a total of 67 layers.
These are surrounded by a scintillation time-of-flight
system and a CsI(Tl) calorimeter with 7800 crystals.
These detector systems are installed inside a
superconducting solenoidal magnet (1.5~T), surrounded by
an iron return yoke instrumented with proportional tube chambers
for muon identification.
 
The $\tau^+\tau^-$ candidate events 
must contain exactly two oppositely charged tracks. 
To suppress beam-gas events, the distance of closest approach
of each track to the interaction point must be within 0.5 cm transverse
to the beam and 5 cm along the beam direction.
We divide each event into two hemispheres (signal and tag), 
each containing one charged track,
by the plane perpendicular to the thrust axis,
which is calculated using both charged tracks
and photons. 
The total invariant mass of the tag hemisphere must be less
than the $\tau$ mass ($M_\tau = 1.777$ GeV/c$^2$)~\cite{pdg}.
Because there is no neutrino in the signal hemisphere
while there is at least one neutrino undetected in the tag hemisphere,
the missing momentum of the event must point toward the tag hemisphere.
To suppress the background from radiative Bhabhas and $\mu$ pairs,
the direction of the missing momentum 
is further required to satisfy $|cos\theta_{missing}| < 0.90$,
where $\theta$ is the polar angle with respect to the beam. 
To reject the background from two-photon interactions,
we require the magnitude of the net transverse momentum vector of each event  
to be greater than 300 MeV/c.

The signal hemisphere must contain an electron or a muon.
The electron candidate must have a shower energy to momentum
ratio in the range, $0.8 < E/p < 1.1$, and have specific ionization
loss within three standard deviations of that expected for an electron.
The muon candidate must penetrate more than
three absorption lengths of material.

Photon candidates are defined as energy clusters in the calorimeter
of at least 60 MeV in the barrel region ($|cos\theta| < 0.80$)
or 100 MeV in the endcap region ($0.80 < |cos\theta| < 0.95$).
We reconstruct $\pi^0$ and $\eta$ mesons using the
$\gamma\gamma$ decay channel.
For the decays involving one meson,
both photons must be in the barrel.
For the decays involving two mesons,
at least one photon from each meson must be in the barrel.
We further require every photon to be separated from the
projection of any charged track by at least 30 cm 
unless its energy is greater than 300 MeV. 
There is no explicit cut on the maximum number of photons
in the signal hemisphere in order to maintain a high
detection efficiency while minimizing the dependence
on the Monte Carlo simulation of electromagnetic showers.
The signal hemisphere may contain
photons not used in the $\pi^0$/$\eta$ reconstruction.
However, photon candidates with energy greater than 300 MeV 
or with a lateral shower profile consistent 
with that expected for a real photon must be used in
the reconstruction.
For the one-meson (two-meson) mode,
events with more than two (four) such
photons in the signal hemisphere are rejected. 

\begin{figure}[p]
\centering
\centerline{\hbox{\psfig{figure=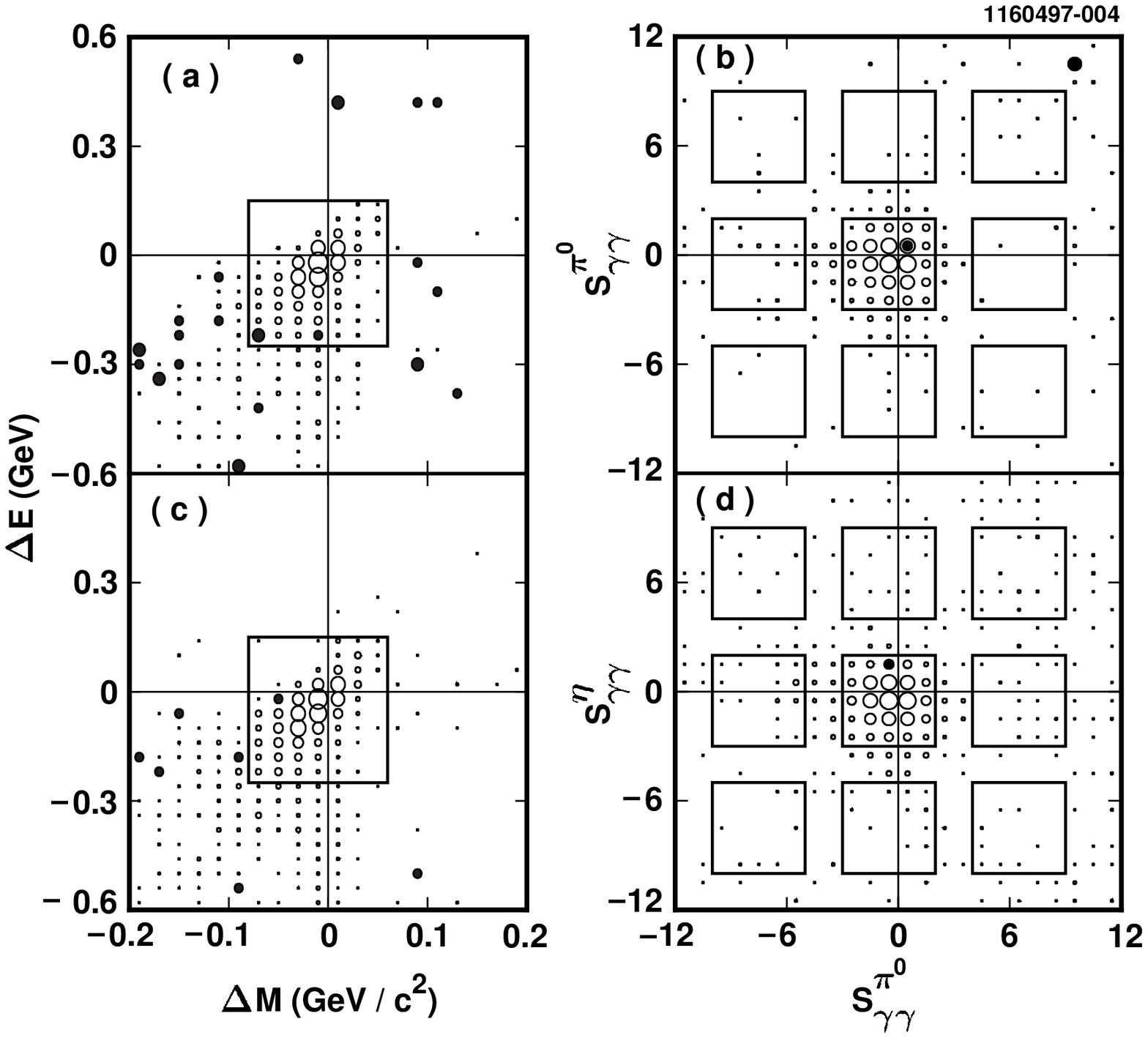}}}
\vspace{0.5in}
\caption{$\Delta E \rm{~vs.~} \Delta M$ and 
$S_{\gamma\gamma} \rm{~vs.~} S_{\gamma\gamma}$ 
distributions of $\tau^- \to \mu^- \pi^0 \pi^0$ (a-b) 
and $\tau^- \to e^- \pi^0 \eta$ (c-d) candidates
in the data and signal Monte Carlo (open circles) samples. 
The $S_{\gamma\gamma} \rm{~vs.~} S_{\gamma\gamma}$ distribution is 
for the center box in the $\Delta E$ vs. $\Delta M$ plane.
The signal, sideband, and corner-band regions 
in the $S_{\gamma\gamma} \rm{~vs.~} S_{\gamma\gamma}$ plane are 
indicated by the 9 boxes. 
The size of the circles is proportional to the number of entries.
The scale for the signal Monte Carlo event distributions is arbitrary.}
\label{fig:mode69}
\end{figure}

To search for neutrinoless $\tau$ decays, we select
$\tau$ candidates with invariant mass $M$ and total energy $E$
within the ranges,
\begin{eqnarray}
\nonumber
-250 < \Delta E = E - \sqrt{s}/2 < 150\ {\rm MeV},\\
\nonumber
-80 < \Delta M = M - M_{\tau} < 60\ {\rm MeV/c}^2.
\end{eqnarray}
\noindent
These requirements correspond approximately to three standard deviation limits,
according to the Monte Carlo simulations (see below).
We then look for $\pi^0$ and $\eta$ candidates using the $\gamma\gamma$ invariant
mass spectrum.
The mass spectrum is expressed in standard deviations
from the nominal $\pi^0$ or $\eta$ mass~\cite{pdg},
\begin{eqnarray}
\nonumber
S_{\gamma\gamma} = ( M_{\gamma\gamma} - M_{\pi^0,\eta} ) / 
\sigma_{\gamma\gamma},
\end{eqnarray}
\noindent where $\sigma_{\gamma\gamma}$ is the mass resolution calculated
from the energy and angular resolution of each photon. 
The $S_{\gamma\gamma}$ distributions may have multiple entries 
due to different combinations of photons in the $\pi^0$/$\eta$ reconstruction.
The signal region is defined as $-3 < S_{\gamma\gamma} < 2$ while
the sideband regions are defined  as $-10 < S_{\gamma\gamma} < -5$ and
$ 4 < S_{\gamma\gamma} < 9$.   
Two signal events satisfy these selection criteria
(see Table~\ref{tab:results}).
The $\Delta E$ vs. $\Delta M$ and $S_{\gamma\gamma}$
distributions of these events are shown in Fig.~\ref{fig:mode69}.

The detection efficiencies ($\epsilon$)
are estimated using a Monte Carlo simulation.
In the Monte Carlo, one $\tau$ lepton decays according to a
two-body (three-body) phase space distribution for the one-meson (two-meson) 
mode of interest and the other $\tau$ lepton decays generically  
according to the KORALB $\tau$ event generator~\cite{KORALB}.
The detector response is simulated using the GEANT program~\cite{GEANT}.
The background from generic $\tau$ decays is estimated using the
KORALB Monte Carlo and the background from hadronic events is
estimated using the Lund Monte Carlo~\cite{Lund}.
The detection efficiencies and background estimates are
summarized in Table~\ref{tab:results}.
Also listed are background estimates based on the numbers of events 
in the sideband ($N_{sb}$) and corner-band ($N_{cb}$), if appropriate, regions
in the $S_{\gamma\gamma}$ distribution,
$N_{bg} = \frac{1}{2} N_{sb} - \frac{1}{4} N_{cb}$,
assuming a linear background distribution.
In the data, $N_{cb}$ is measured to be zero for all six two-meson decays. 
Because of the paucity of events, rather than comparing the number of
events observed to the expected background in each individual mode, we
will sum over all the modes for comparison.
The two events observed is somewhat higher than the 0.5
events expected from the sideband technique but consistent
with the 1.5 events estimated by the Monte Carlo simulations.
There is therefore no evidence for a signal.
The background estimated using the sideband technique is used
to compute the upper limit on the signal.

\begin{table}[t]
\begin{center}
\caption[]{Summary of detection efficiencies, signal, backgrounds,
90\% C.L. upper limits on the signal (see text) and branching fractions.}
\vspace{0.1in}
\label{tab:results}
\begin{tabular}{ccccccccccc}
Mode & $\epsilon(\%)$ & $N_{ob}$ & $N_{bg}$ 
& $N_{bg}^{\tau MC}$ & $N_{bg}^{q \bar q\ MC}$ & $\lambda_0$
& $\lambda_G$ & $\lambda_P$ & $\lambda$ & ${\cal B}(10^{-6})$ \\
\hline                                                  
$e^- \pi^0$         &8.71 & &   &   &   & 2.30 & 2.32 & 2.30 & 2.32 & 3.72 \\
$\mu^- \pi^0$       &8.08 & &0.5&   &   & 2.30 & 2.32 & 2.30 & 2.32 & 4.01 \\
$e^- \eta$          &9.97 & &   &   &   & 2.30 & 2.32 & 2.30 & 2.32 & 8.19 \\
$\mu^- \eta$        &8.49 & &   &   &   & 2.30 & 2.32 & 2.30 & 2.32 & 9.62 \\
$e^- \pi^0 \pi^0$   &5.12 & &   &   &0.4& 2.30 & 2.34 & 2.30 & 2.34 & 6.47 \\
$\mu^- \pi^0 \pi^0$ &3.71 &1&   &0.5&   & 3.89 & 3.97 & 3.61 & 3.68 & 14.0 \\
$e^- \eta \eta$     &6.09 & &   &   &   & 2.30 & 2.34 & 2.30 & 2.34 & 34.5 \\
$\mu^- \eta \eta$   &3.48 & &   &   &   & 2.30 & 2.34 & 2.30 & 2.34 & 60.2 \\
$e^- \pi^0 \eta$    &5.52 &1&   &   &   & 3.89 & 3.96 & 3.61 & 3.68 & 23.8 \\
$\mu^- \pi^0 \eta$  &3.73 & &   &0.2&0.4& 2.30 & 2.34 & 2.30 & 2.34 & 22.4 \\
\end{tabular}
\end{center}
\end{table}

The upper limit on the branching fraction is related to the
upper limit $\lambda$ on the signal by
\begin{eqnarray}
\nonumber
{\cal B} =
\frac { \lambda}{ 2 \epsilon N_{\tau\tau} {\cal B}_1 {\cal B}^m_{\pi^0}
{\cal B}^n_{\eta}}
\end{eqnarray}
where ${\cal B}_1$ is the inclusive 1-prong
branching fraction~\cite{pdg},
${\cal B}_{\pi^0}$ (${\cal B}_{\eta}$) is the branching fraction~\cite{pdg} for
$\pi^0 \to \gamma\gamma$ ($\eta \to \gamma\gamma$), and $m$ ($n$)
is the number of $\pi^0$ ($\eta$) mesons in the final state.
The 90\% confidence level upper limits on the signal
are summarized in Table~\ref{tab:results}.
We calculate the upper limit $\lambda$ using a Monte Carlo technique,
which incorporates both the Poisson statistics of the signal and the
systematic errors.
The systematic errors include the statistical uncertainty in the
background estimate due to limited statistics in the sideband
(and corner-band if appropriate) regions.
This statistical uncertainty is incorporated using Poisson statistics.
All other sources of systematic errors are incorporated
using Gaussian statistics.  
These include the uncertainties in the $\tau^+\tau^-$ cross section (1.0\%),
luminosity (1.0\%), track reconstruction (3.0\%),
lepton identification (1.5\% for $e$ and 4.0\% for $\mu$),  
$\pi^0$ or $\eta$ meson reconstruction (5.0\% per meson),
branching fraction of $\eta\to\gamma\gamma$
(0.8\%)~\cite{pdg}, and
detection efficiencies due to limited
Monte Carlo statistics (2-3\% for the one-meson
modes and 3-4\% for the two-meson modes).
These uncertainties are added in quadrature in computing $\lambda$.
For comparison, we also list the upper limits $\lambda_0$, $\lambda_G$ and $\lambda_P$.
$\lambda_0$ is calculated using only Poisson statistics for the signal,
$\lambda_G$ includes all the systematic errors except the statistical
uncertainty in the background estimate, and $\lambda_P$ includes only
the statistical error in the background estimate.
$\lambda_G$ is larger than $\lambda_0$ as expected.
However, $\lambda_P$ is smaller than $\lambda_0$
when the observed number of events is non-zero and
the estimated background is zero.
This is not unexpected because in the calculation of $\lambda_P$ 
we allow for the estimated zero background to fluctuate up, 
in contrast to $\lambda_0$ 
in which the background is estimated to be zero with no uncertainty.

The upper limits on the branching fractions for the modes
involving one neutral meson are significantly 
more stringent than the published results~\cite{pdg}.
There are no previous limits for the modes involving two neutral mesons.
The limits for the $\pi^0\pi^0$ modes are comparable
with the limits for the $\pi^+\pi^-$ modes~\cite{pdg}.
In the model of Gonzalez-Garcia and Valle~\cite{Valle}, the limit
on $\tau^- \to e^- \pi^0$ extends the heavy lepton mass vs. mixing region
previously excluded from other measurements:
neutral leptons with mass greater than 6-10 TeV/c$^2$ for mixing with
the third generation in the range 0.05-0.02 are now excluded~\cite{Garcia}.

We gratefully acknowledge the effort of the CESR staff in providing us with
excellent luminosity and running conditions.
This work was supported by 
the National Science Foundation,
the U.S. Department of Energy,
the Heisenberg Foundation,  
the Alexander von Humboldt Stiftung,
Research Corporation,
the Natural Sciences and Engineering Research Council of Canada,
and the A.P. Sloan Foundation.


\newpage

\begin{thebibliography}{99}
\bibitem{light}
Throughout this paper, the charge conjugate state is implied
and $\ell$ represents either an electron or a muon.
\end{thebibliography}
\end{document}